\documentclass[sigconf,screen,authorversion]{acmart}

\AtBeginDocument{%
  }

\setcopyright{none}
\copyrightyear{2024}
\acmYear{2024}
\acmDOI{}

\acmConference[FAccTRec '24]{The 7th FAccTRec Workshop: Responsible Recommendation}{October 14, 2024}{}
\acmISBN{978-1-4503-XXXX-X/18/06}

\begin{document}


\title[An Intervention Tool for End-User Algorithm Audits]{Designing an Intervention Tool for End-User Algorithm Audits in Personalized Recommendation Systems}

\author{Qunfang Wu}
\authornote{Both authors contributed equally to this research.}
\affiliation{%
  \institution{Harvard University}
  \city{Cambridge}
  \state{Massachusetts}
  \country{USA}
}
\email{qunfangwu@fas.harvard.edu}
\author{Lu Xian}\authornotemark[1]
\affiliation{%
  \institution{University of Michigan}
  \city{Ann Arbor}
  \state{Michigan}
  \country{USA}
}
\email{xianl@umich.edu}



\renewcommand{\shortauthors}{Wu and Xian}




\begin{abstract}
As algorithms increasingly shape user experiences on personalized recommendation platforms, there is a growing need for tools that empower end users to audit these algorithms for potential bias and harms. This paper introduces a novel intervention tool, \textit{MapMyFeed}, designed to support everyday user audits. The tool addresses key challenges associated with user-driven algorithm audits, such as low algorithm literacy, unstructured audit paths, and the presence of noise. MapMyFeed assists users by offering guiding prompts, tracking audit paths via a browser extension, and visualizing audit results through a live dashboard. The tool will not only foster users’ algorithmic literacy and awareness but also enhance more transparent and fair recommendation systems.
\end{abstract}

\maketitle

\section{Introduction}

Algorithm audits examine the performance and effects of algorithmic systems on people especially when the internals of these systems remain opaque \cite{metaxa_auditing_2021}. They serve as tools for accountability that help mitigate undesirable consequences of algorithms \cite{metaxa_auditing_2021}. 
The approach to conducting such audits is shaped by various factors including the auditors (who execute audits), stakeholders (who are affected by or benefit from audits), and evaluation metrics (what aspects are audited). Auditors can range from internal teams within companies to firms specializing in audits, as well as independent organizations and individuals \cite{costanza2022audits}.
Stakeholders range from end-users, who interact with technologies like recommendation systems \cite{ribeiro2020auditing} and facial recognition \cite{buolamwini2018gender}, to regulators and algorithm vendors, each with distinct key interests such as personalized relevance of the recommendations, user privacy, policy-making, and risk management, respectively \cite{brown2021algorithm}. The choice of evaluation metrics, from discrimination/bias to efficiency or transparency \cite{bandy2021problematic}, further influences audit methodologies~\cite{brown2021algorithm, costanza2022audits}. 
As such, developing effective algorithm audits requires an understanding of the interplay among auditors, the identification of stakeholders, and evaluation metrics~\cite{ayling2022putting, sandvig2014auditing, brown2021algorithm}. In this paper, we focus on a particular auditing approach, \textit{end-user audits}, in the context of social media recommendation algorithms, and propose an intervention tool for this approach. Building on prior literature, our proposed intervention positions end users as both auditors and key stakeholders, and develops user-centered metrics to assist their audits.

End-user audits leverage users' everyday interactions with algorithmic systems to surface harmful behaviors of the systems that are often hard to detect through formal audits conducted outside the everyday use context \cite{seaver2017algorithms, friedman1996bias}. Initially proposed by \citet{shen_everyday_2021}, the notion of user everyday audit is borrowed from Suchman's ~\cite{suchman1987plans} situational actions, where users adapt and decide their actions based on immediate circumstances rather than strictly following a pre-defined plan or script. Users often face unexpected or problematic algorithm recommendations through everyday interactions with algorithmic systems. In response, they adapt their actions dynamically, leveraging the context of the situation to better understand and effectively respond to what they have faced \cite{suchman1987plans, cen_user_2023, lee_algorithmic_2022}.

Different from formal, systemic audits that often consider regulators and society at large as key stakeholders \cite{brown2021algorithm,metaxa_auditing_2021}, everyday audits consider users as key stakeholders and empower them to understand and manage these algorithms.
Prior work has underscored the value of centering around user interactions and experiences with the platforms in understanding the interdependent relations between users, algorithms, and social media platforms \cite[e.g.,][]{cotter2019playing} and in empowering resistance against algorithmic biases and harms \cite[e.g.,][]{velkova2021algorithmic, karizat2021algorithmic}. Scholars have also investigated users' algorithm training strategies~\cite{cen_measuring_2024,burrell_when_2019,haupt_recommending_2023} and user audit models~\cite{devos_toward_2022,lam_end-user_2022,li_participation_2023} across different algorithmic systems. While this prior scholarship provides a foundation for conducting user audits, little work has discussed and designed interventions or educational tools to address the challenges of user everyday audits and assist users in performing these audits. Our contributions include designing a new tool for user everyday audits and examining intervention mechanisms (e.g., guiding prompts, audit visualization) to better support user audits. The proposed tool aims to improve users’ literacy and awareness of algorithmic impacts while promoting more transparent and fair recommendations. In this position paper, we first discuss the challenges of user everyday audits and then propose an intervention tool that addresses some, if not all, of these challenges for personalized recommendation systems and briefly delineates its functions. We conclude by discussing the future work. 

\section{Challenges of User Everyday Audits}
We categorize the challenges into four aspects: (1) lack of auditor expertise for novice and low-literacy users, (2) unstructured audit paths, (3) noise in audit performance, and (4) obstacles to transitioning to collective audits.

\subsection{Novice and Low-Literacy Users}

End-users vary in their algorithm literacy and expertise. While many user audit cases show that users have some algorithm expertise, others reveal that those with little or no algorithm knowledge need support from researchers or their technically savvy peers~\cite{lei2021delivering,qadri2021delivery}. A user algorithm audit usually starts with problem initiation~\cite{shen_everyday_2021}. Users, especially those new to or with low algorithm literacy, may initiate audits based on vague perceptions and speculations. Users’ experiences and knowledge also affect their ability to identify algorithm biases. As~\citet{devos_toward_2022} found, people’s experiences determine the types of bias they recognize; they often relied on second-hand knowledge and exhibited their own cognitive biases. Unlike crowd-sourced audits, where users are passive data contributors, everyday audits require users to lead their audits as “investigators”~\cite{lam_end-user_2022}. Therefore, intervention tools should offer necessary knowledge and prompts for users at different levels of algorithm literacy to develop audit goals and strategies. 

\subsection{Unstructured Audit Paths}

User algorithm audits benefit from the user’s everyday use of the system, making the audit process organic~\cite{shen_everyday_2021}. This organic nature is reflected in the entire audit process: a user may start an audit spontaneously, conduct the audit intermittently, and even carry out multiple audits simultaneously. Thus, user everyday audits follow a “non-linear” path and are unstructured~\cite{shen_everyday_2021}. This process differs from prior approaches in explainable AI or interactive recommendation systems, where users were guided by a fixed or semi-fixed procedure to understand the internal workings of a system and ways to achieve a desirable outcome.
Prior research has attempted to understand the process of users conducting algorithm audits~\cite{devos_toward_2022, shen_everyday_2021}. For example, \citet{devos_toward_2022} found that users followed a process of “search inspiration,” “sensemaking,” and “remediation” while identifying harmful algorithms. However, to guide fluid forms of user audits, the steps and process they theorized about need to be translated into adaptable and feasible guidance and supported by tools. An intervention tool can support unstructured audits by recording user actions during the audit process and organizing them into more structured formats, such as topics and timelines, for easier sensemaking.

Another issue with the unstructured nature of audits is users’ cognitive load. User audits often occur during routine system usage, where users may not consistently focus on auditing tasks. Additionally, some algorithmic issues require long-term observations and evaluations. Therefore, an intervention tool should reduce cognitive load by internally reminding users of their audit tasks and recording audit actions and outcomes over extended periods.

\subsection{Noise}

User audits are less controlled or systematic compared to internal or third-party audits, which can lead to potential noise and errors. Identifying the signal versus noise is a well-known challenge for user- and crowd-driven approaches~\cite{devos_toward_2022}. Additionally, various situational factors might influence the outcomes of user audits. These factors include the user’s environment, the context of algorithm use, and individual biases. Understanding these factors is crucial for identifying and mitigating noise or errors~\cite{devos_toward_2022}. \citet{casper_black-box_2024} discussed that black-box audits have inherent limitations that can lead to noise. Many black-box methods for explaining model decisions are unreliable because they often fail to accurately identify causal relationships between input features and outputs. For example, users might incorrectly attribute an algorithm’s decision to irrelevant features, misleading engineers about the true cause of the bias. To ensure the accuracy and integrity of the audit, the tool needs to manage noise effectively. We also acknowledge that algorithmic biases are often first identified through isolated incidents or personal anecdotes reported by users. Although these individual reports may not reflect systemic issues, evidence of harm from even a single user is valuable, regardless of its replicability or prevalence across the platform. These reports can reveal issues impacting minority groups or edge cases that might otherwise remain hidden. Therefore, the intervention tool should not minimize ``noise'' to enhance scientific robustness but utilize it to identify and address specific harms experienced by underrepresented or marginalized groups.

\subsection{Transitioning to Collective Audits}

Individual user audits often transition into collective actions. As previously mentioned, algorithm audits extend beyond individual test cases to make broader declarations about the system as a whole~\cite{metaxa_auditing_2021}. However, user audit is limited in developing a comprehensive understanding of the system~\cite{casper_black-box_2024}. In contrast, collective audits can more easily accumulate evidence and validate individual findings. 

Supporting collective audits requires addressing issues throughout different audit stages.
Users first need to understand the audit task, a process known as collective sensemaking ~\cite{devos_toward_2022,shen_everyday_2021}. However, \citet{devos_toward_2022} found that the lack of contextual information made it difficult for the group to differentiate between various explanations for observed behaviors based on user reports.
For example, explanations could vary among users from different demographics. Without this demographic context, the group found it difficult to understand these different explanations. They suggested that design solutions could allow for follow-up questions to those reporting their observations or provide reminders for auditors to consider missing information~\cite{devos_toward_2022}. 
Other challenges for collective audits include role assignment~\cite{li_participation_2023}, data sharing~\cite{lam_end-user_2022}, and task coordination~\cite{shen_everyday_2021}. 
Such nuanced collective audits necessitate robust support mechanisms. 
Our intervention tool primarily focuses on individual user audits but is also designed to be adaptable for potential collective audits.

\section{Designing a User Audit Intervention Tool for Personalized Recommendation Systems}

In response to these challenges, we propose a user audit intervention tool design for personalized recommendation systems. We consider a common scenario: users' interactions with the system depend not only on the content recommended to them but on their attempts to shape future recommendations. For example, on personalized recommendation platforms, users often leverage their understanding of important engagement functions, including likes, comments, and dwell time, to manipulate the algorithm \cite{cen2024measuring}. 

Our tool, \textit{MapMyFeed}, assists everyday users in navigating the auditing process through three key features: (1) it offers guiding prompts that help users initiate and conduct audits, (2) it uses a browser extension to track the audit path, and (3) it employs a live dashboard to visualize and communicate the audit results. 

The guiding prompts in our tool are crafted to assist novice and low-literacy users by equipping them with the thinking tools needed to navigate the auditing process. Drawing on explainable AI and algorithmic audit literature \cite[e.g.,][]{liao2020questioning, liao2021question, shen_everyday_2021, brown_algorithm_2021, ehsan2021operationalizing, chromik2021human}, we have developed prompts that encourage users to engage critically with the system. The semi-structured guiding prompts offer initial context for users to start an audit with specific, self-defined goals, after which they are encouraged to proceed with organic interaction and exploration. 
They prompt users to consider the relevance of recommendations (system performance), explore hypothetical changes in system outputs if certain features are altered (``what if'' scenarios), and make preliminary hypotheses about how the system functions and the impact of its features (how). This semi-structured approach not only aids users in establishing audit objectives and metrics and understanding the steps of effective audits, but also allows them the flexibility to adhere to their preferred practices or explore new routines in their interactions with the recommender system.

The recording of audit paths meticulously logs users’ exposure to and interactions with the system during an audit, using a browser extension. This extension tracks user behaviors---for instance, on a social media platform---including viewing, clicking, liking, saving, sharing, and following. 
Such historical data provides the context for users to understand their engagement with the system \cite{zannettou2024analyzing} and how their engagement shapes audit outcomes. Our semi-structured approach to audit goal setting, combined with the documentation of audit paths, helps address the issue of non-linear and unstructured audit paths in user everyday audits. 
The collected user data will be anonymized before further analysis and presentation; user data will be used for research purposes only. 

The live dashboard offers users reports summarizing their audits on the platform. These reports detail their interactions and audit outcomes, encourage them to reflect on their audit objectives, and enable them to review and compare their results with those of others in an aggregated format. This feature facilitates a deeper analysis and provides a broader context for understanding each user's audit results. The reports filter out noise to ensure the accuracy of audit outcomes. 

The tool offers educational resources to guide them in understanding the mechanisms of algorithmic recommendations. This proactive involvement allows users to identify and report problematic or biased algorithmic behaviors, cultivating their awareness of fairness issues in recommendation systems. 
By fostering algorithmic literacy, the tool not only educates users but also helps make the internal workings of the system and its impacts legible to users through their daily interactions. As an accountability mechanism, our tool supports the development of more equitable digital environments.

\section{Future Work}
In the next phase of this research, we will focus on further developing and refining the proposed tool. We will conduct controlled lab experiments to evaluate how effectively the tool supports users in addressing key challenges, such as unstructured audit paths, noise management, and algorithmic literacy. Specifically, to assess the proposed intervention mechanisms, we will conduct user studies, including surveys, interviews, and think-aloud sessions, to capture users' experiences with features like guiding prompts and audit outcome visualization. Through iterative design cycles, feedback gathered from these studies will guide ongoing improvements to the tool. In the long term, we will conduct longitudinal field studies to evaluate the tool's performance in real-world environments, tracking users' audit behaviors over time to assess how the tool improves user algorithmic literacy and its ability to surface algorithmic biases.

\bibliographystyle{ACM-Reference-Format}
\bibliography{references,references-local}

\end{document}